\documentclass[eqsecnum,nofootinbib,11pt]{revtex4}
\usepackage{amsmath}
   \usepackage{bm}
\usepackage{amsthm,amscd}
\usepackage{mathrsfs}
\usepackage{verbatim}
\usepackage{amsfonts,amsmath,latexsym,amssymb,txfonts,bm}
\usepackage[american]{babel}
\usepackage{dsfont}
\usepackage[dvips]{graphicx}

\newcommand\bea{\begin{eqnarray}}
\newcommand\eea{\end{eqnarray}}

\newtheorem{lemma}{Lemma}[]

\begin{document}
\title{On the Hurwitz Zeta Function of Imaginary Second Argument}
\author{
Guglielmo Fucci\footnote{Electronic address: Guglielmo\textunderscore Fucci@Baylor.edu}
\thanks{Electronic address: gfucci@nmt.edu}}
\affiliation{Department of Mathematics, Baylor University, Waco, TX 76798 USA
}
\date{\today}
\vspace{2cm}
\begin{abstract}

In this work we exploit Jonqui\`{e}re's formula relating the Hurwitz zeta function to a linear combination of
polylogarithmic functions in order to evaluate the real and imaginary part of $\zeta_{H}(s,ia)$ and its first derivative with respect to the first argument $s$.
In particular, we obtain expressions for the real and imaginary party of $\zeta_{H}(s,i a)$ and its derivative for $s=m$ with $m\in\mathbb{Z}\backslash\{1\}$ involving
simpler transcendental functions.

\end{abstract}
\maketitle

\section{Introduction}

The Hurwitz zeta function, which is a generalization of the Riemann zeta function, is defined for $\Re(s)>1$ and $\Re(a)\neq-m$, with $m\in\mathbb{N}_{0}$, through the series
\begin{equation}\label{0}
  \zeta_{H}(s,a)=\sum_{n=0}^{\infty}\frac{1}{(n+a)^{s}}\;,
\end{equation}
and can be analytically continued in a unique way to a meromorphic function in the entire complex plane possessing only a simple pole with residue
$1$ at the point $s=1$. This higher transcendental function is, amongst others, of fundamental importance in a wide range of mathematical areas stemming from number theory to analysis \cite{apostol95}. In physics, its importance lies in the regularization procedures used in areas such as quantum field theory at zero and finite temperature, string theory, etc. (see e.g. \cite{bytsenko03,elizalde94,elizalde,kirsten01}). The analytic continuation of the Hurwitz zeta function in the semi-plane $\Re(s)<1$ is a well known subject which can be found in a variety of classic texts on special functions. It is also worth mentioning that the first and higher derivatives of the Hurwitz zeta function have been analyzed in \cite{elizalde86,elizalde93,elizalde95,miller98}.

The relevance of the Hurwitz zeta function of imaginary second argument is related to the phenomenon, termed Schwinger mechanism, of pair production under the influence of a strong
electric field \cite{schwinger51}.

It was shown in \cite{schwinger51} that the one-loop Lagrangian density for a massive field under the influence of an electric field in Minkowski spacetime becomes a meromorphic
function with isolated simple poles located on the real axis. The effective action is then obtained by integrating the Lagrangian density and avoiding these poles.
This procedure generates an imaginary part in the effective action which is interpreted as pair production rate. Let us point out that this procedure has been exploited on general manifolds without boundary to obtain the gravitational corrections to the Schwinger mechanism \cite{avramidi09,fucci09,fucci10}.

By utilizing zeta function regularization techniques, it was shown in \cite{blau91} that the one-loop effective action is expressed in terms of $\zeta_{H}(s,i x)$ and $\zeta'_{H}(s,i x)$ with  the dimensionless constant $x=m^{2}/2|E|$ where $m$ represents the mass of the field under consideration, $|E|$ the strength of the electric field and the prime denotes differentiation with respect to the variable $s$. In particular, the rate of creation of pairs under the influence of a strong electric field on a $D$-dimensional Minkowski spacetime was explicitly computed and the results
written in terms of the imaginary part of $\zeta_{H}(s,i x)$ and $\zeta'_{H}(s,i x)$ \cite{blau91}.
In \cite{adesi02}, by utilizing a formula regarding the analytic continuation of the first derivative of the Hurwitz zeta function of imaginary second argument, the authors were able to prove that the results obtained in \cite{blau91} indeed coincide with the pair production rate obtained by Schwinger in \cite{schwinger51}. However, in \cite{adesi02}, the results were limited to explicit expressions for the real part of $\zeta'_{H}(s,i x)$ at even negative integers and its imaginary part at odd negative integers which are the relevant ones for the purpose of analyzing the pair production rate of massive spinor fields in $4$-dimensional Minkowski spacetime.

The aim of this work is to extend the results obtained in \cite{adesi02} in order to include expressions for the
real and imaginary part of the Hurwitz zeta function of imaginary second argument, and its derivative, for all integers $m\in\mathbb{Z}\backslash\{1\}$. These results could be immediately applied, for instance, to the computation of the pair production rate of massive bosons and spinors in higher dimensional Minkowski spacetimes.

In the rest of this work, we will be mainly interested in the range $x\in(0,1)$ which is the relevant one for the strong electric field regime, namely $|E|\gg m^{2}$.

The outline of the paper is as follows. We will utilize Jonqui\`{e}re's formula relating the Hurwitz zeta function $\zeta_{H}(s,a)$ to a linear combination of
polylogarithmic functions in order to extract the real and imaginary part of $\zeta_{H}(s,i x)$. We will then find expressions for $\zeta_{H}(s,i x)$ and its first derivative
at $s=-m$, with $m\in\mathbb{N}_{0}$. In the second part of this work we will present formulas for $\zeta_{H}(s,ix)$ and $\zeta'_{H}(s,ix)$ valid for all positive integers $s=m$, with $m\in\mathbb{N}\backslash\{1\}$.

\section{Jonqui\`{e}re's representation of the Hurwitz zeta function}\label{sec2}

The starting point of our analysis is the following representation of the Hurwitz zeta function valid for $0\leq\Re(z)<1$ and $\Im(z)\geq 0$ \cite{jonquiere89}
\begin{equation}\label{1}
  i^{-s}\textrm{Li}_{s}\left(e^{2\pi i z}\right)+i^{s}\textrm{Li}_{s}\left(e^{-2\pi i z}\right)=\frac{(2\pi)^{s}}{\Gamma(s)}\zeta_{H}(1-s,z)\;,
\end{equation}
where $\textrm{Li}_{s}(w)$ represents the polylogarithmic function and $\Gamma(s)$ the Euler gamma function. For the purpose of this work, we will set $\Re(z)=0$ and
$\Im(z)=x\in(0,1)$ and hence the formula (\ref{1}) specialized to our case becomes
\begin{equation}\label{2}
  \zeta_{H}(s, i x)=\frac{i\Gamma(1-s)}{(2\pi)^{1-s}}\left[i^{-s}\textrm{Li}_{1-s}\left(e^{2\pi x}\right)-i^{s}\textrm{Li}_{1-s}\left(e^{-2\pi x}\right)\right] \;,
\end{equation}
which is well defined for $\Re(s)<1$ and can be extended to the entire complex plane by analytic continuation. In fact for $s=1+\varepsilon$, formula
(\ref{2}) reduces, as $\varepsilon\to 0$, to
\begin{equation}\label{3}
  \zeta_{H}(1+\varepsilon, i x)=-\frac{1}{\varepsilon}\left[\textrm{Li}_{0}\left(e^{2\pi x}\right)+\textrm{Li}_{0}\left(e^{-2\pi x}\right)\right]+O(1)\;,
\end{equation}
and by exploiting the fact that $\textrm{Li}_{0}\left(w\right)=w/(1-w)$ we recover the simple pole of the Hurwitz zeta function with the correct residue.
On the other hand, for $s=n+1$ with $n\in\mathbb{N}_{0}$, the simple pole of the gamma function in (\ref{2}) does not translate into a pole in $\zeta_{H}(s,i x)$ as one should expect
thanks to the following relation satisfied by the polylogarithmic functions \cite{erdelyi53}
\begin{equation}\label{4}
  \textrm{Li}_{-n}\left(w\right)+(-1)^{n}\textrm{Li}_{-n}\left(w^{-1}\right)=0\;.
\end{equation}

By rewriting $i^{\pm s}$ in terms of trigonometric functions, equation (\ref{2}) can be cast into the form
\begin{eqnarray}\label{5}
  \zeta_{H}(s, i x)&=&\frac{\Gamma(1-s)}{(2\pi)^{1-s}}\sin\left(\frac{\pi}{2}s\right)F(s,x)+i\frac{\Gamma(1-s)}{(2\pi)^{1-s}}\cos\left(\frac{\pi}{2}s\right)G(s,x)\;,
\end{eqnarray}
where we have defined, for convenience, the following functions
\begin{equation}\label{5a}
  F(s,x)=\textrm{Li}_{1-s}\left(e^{2\pi x}\right)+\textrm{Li}_{1-s}\left(e^{-2\pi x}\right)\quad\textrm{and}\quad  G(s,x)=\textrm{Li}_{1-s}\left(e^{2\pi x}\right)-\textrm{Li}_{1-s}\left(e^{-2\pi x}\right)=F(s,x)-2\textrm{Li}_{1-s}\left(e^{-2\pi x}\right)\;.
\end{equation}
The representation (\ref{5}) is particularly suitable for extracting the real and imaginary part of the Hurwitz zeta function $\zeta_{H}(s,ix)$ once the real and imaginary part of $F(s,x)$ and $G(s,x)$ are known.

For this purpose, we notice that for $x<1$, which is within the assumptions of our work, the function $\textrm{Li}_{1-s}(e^{-2\pi x})$ is real when $s$ and $x$ assume real values,
while $\textrm{Li}_{1-s}(e^{2\pi x})$ is a complex function for $\Re(s)<1$ and $x$ real, and is a real function for $\Re(s)\geq 1$ with $x$ real. This particular behavior
suggests us to distinguish between two different cases:

\begin{paragraph}{Semi-plane $\Re(s)<1$.} In this case, $\textrm{Li}_{1-s}(e^{2\pi x})$ is complex for real $x$.
However, the real and imaginary part of $\textrm{Li}_{1-s}(e^{2\pi x})$
can be identified by exploiting the following series representation valid for $s\in\mathbb{C}/\mathbb{N}$ and $|x|<1$ \cite{erdelyi53}
\begin{equation}\label{6}
  \textrm{Li}_{s}(e^{2\pi x})=\Gamma(1-s)(-2\pi x)^{s-1}+\sum_{k=0}^{\infty}\frac{\zeta_{R}(s-k)}{k!}(2\pi x)^{k}\;,
\end{equation}
where $\zeta_{R}(s)$ represents the Riemann zeta function.
The only contribution to the imaginary part of (\ref{6}) comes from the first term, and it is not very difficult
to obtain the decomposition
\begin{eqnarray}
  \Re F(s,x)&=&\frac{\pi}{\tan(\pi s)}\frac{(2\pi x)^{-s}}{\Gamma(1-s)}+\sum_{k=0}^{\infty}\frac{\zeta_{R}(1-s-k)}{k!}(2\pi x)^{k}+\textrm{Li}_{1-s}\left(e^{-2\pi x}\right)\;,\label{7a}\\
  \Im F(s,x)&=&-\pi\frac{(2\pi x)^{-s}}{\Gamma(1-s)}\;.\label{7}
  \end{eqnarray}
In addition, from the relation (\ref{5a}), we have
\begin{equation}\label{7b}
  \Re G(s,x)=\Re F(s,x)-2\textrm{Li}_{1-s}\left(e^{-2\pi x}\right)\;,\quad \Im G(s,x)=\Im F(s,x)\;.
\end{equation}

The results derived above can be utilized, together with the general formula (\ref{5}), in order to obtain expressions for the real and imaginary part of $\zeta_{H}(s,ix)$, namely
\begin{eqnarray}
  \Re\zeta_{H}(s,ix)&=&\frac{\Gamma(1-s)}{(2\pi)^{1-s}}\left[\sin\left(\frac{\pi}{2}s\right)\Re F(s,x)-\cos\left(\frac{\pi}{2}s\right)\Im G(s,x)\right]\;,\label{8}\\
   \Im\zeta_{H}(s,ix)&=&\frac{\Gamma(1-s)}{(2\pi)^{1-s}}\left[\cos\left(\frac{\pi}{2}s\right)\Re G(s,x)+\sin\left(\frac{\pi}{2}s\right)\Im F(s,x)\right]\;.\label{8a}
\end{eqnarray}

Starting from the formulas above we can write down an expression for the real and imaginary part of $\zeta'_{H}(s,ix)$. In fact, by differentiating
(\ref{8}) and (\ref{8a}) and by using the explicit form of $\Im F(s,x)$ and $\Im G(s,x)$ in (\ref{7}) one gets
\begin{eqnarray}\label{22}
  \Re\zeta'_{H}(s,ix)&=&\frac{\Gamma(1-s)}{(2\pi)^{1-s}}\sin\left(\frac{\pi}{2}s\right)\big\{\Re F(s,x)\left[\ln 2\pi-\Psi(1-s)\right]+\left(\Re F\right)'(s,x)\big\}-\frac{\pi}{4}\sin\left(\frac{\pi}{2}s\right)x^{-s}\nonumber\\
  &+&\frac{\Gamma(1-s)}{(2\pi)^{1-s}}\cos\left(\frac{\pi}{2}s\right)\left\{\frac{\pi}{2}\Re F(s,x)-\frac{(2\pi)^{1-s}}{2\Gamma(1-s)}x^{-s}\ln x\right\}\;,
\end{eqnarray}
and a similar expression for the imaginary part
\begin{eqnarray}\label{22a}
  \Im\zeta'_{H}(s,ix)&=&\frac{\Gamma(1-s)}{(2\pi)^{1-s}}\cos\left(\frac{\pi}{2}s\right)\big\{\Re G(s,x)\left[\ln 2\pi-\Psi(1-s)\right]+\left(\Re G\right)'(s,x)\big\}-\frac{\pi}{4}\cos\left(\frac{\pi}{2}s\right)x^{-s}\nonumber\\
  &-&\frac{\Gamma(1-s)}{(2\pi)^{1-s}}\sin\left(\frac{\pi}{2}s\right)\left\{\frac{\pi}{2}\Re G(s,x)-\frac{(2\pi)^{1-s}}{2\Gamma(1-s)}x^{-s}\ln x\right\}\;,
\end{eqnarray}
where $\Psi(n)$ represents the logarithmic derivative of the gamma function and the prime denotes differentiation with respect to $s$.
The derivative $\left(\Re F\right)'(s,x)$ in (\ref{22}) can be found to have the form
\begin{eqnarray}\label{9}
  \left(\Re F\right)'(s,x)&=&-\frac{\pi}{\tan\pi s}\frac{(2\pi x)^{-s}}{\Gamma(1-s)}\left[\frac{2\pi}{\sin2\pi s}+\ln2\pi x-\Psi(1-s)\right]\nonumber\\
  &-&\sum_{k=0}^{\infty}\frac{\zeta'_{R}(1-s-k)}{k!}(2\pi x)^{k}-\textrm{Li}_{1-s}^{(1)}\left(e^{-2\pi x}\right)\;,
\end{eqnarray}
where the apex in the polylogarithmic function indicates differentiation with respect to its order. Obviously, a similar formula for $\left(\Re G\right)'(s,x)$
can be obtained by differentiating (\ref{7b}) and by using (\ref{9}).
\end{paragraph}
\begin{paragraph}{Semi-plane $\Re(s)\geq 1$.} In this situation, the function $\textrm{Li}_{1-s}(e^{2\pi x})$ is real when the variable $x$ is real and
the representation (\ref{5}) is sufficient in order to immediately extract the real and imaginary part of $\zeta_{H}(s,ix)$. An expression for the first derivative which is suitable in the region $\Re(s)\geq 1$ is obtained by differentiating (\ref{5}), i.e.
\begin{eqnarray}\label{23}
  \zeta'_{H}(s,ix)&=&\frac{\Gamma(1-s)}{(2\pi)^{1-s}}\sin\left(\frac{\pi}{2}s\right)\left[F(s,x)\left(\ln 2\pi-\Psi(1-s)\right)+F'(s,x)\right]+\frac{\pi}{2}\frac{\Gamma(1-s)}{(2\pi)^{1-s}}\cos\left(\frac{\pi}{2}s\right)F(s,x)\nonumber\\
  &+&i\frac{\Gamma(1-s)}{(2\pi)^{1-s}}\cos\left(\frac{\pi}{2}s\right)\left[G(s,x)\left(\ln 2\pi-\Psi(1-s)\right)+G'(s,x)\right]-i\frac{\pi}{2}\frac{\Gamma(1-s)}{(2\pi)^{1-s}}\sin\left(\frac{\pi}{2}s\right)G(s,x)\;.\nonumber\\
\end{eqnarray}
The above expression will be the starting point for the computation of the real and imaginary part of $\zeta'_{H}(s,ix)$ when
$s$ is a positive integer. A detailed analysis of this case is presented in section \ref{sec4}.

\end{paragraph}

The formulas that we have obtained in this section are valid, in their respective ranges of $s$, for $|x|<1$ and are written in terms of simpler transcendental functions, namely the Riemann and
the polylogarithmic function and their derivative. The series representations that we have exploited are quickly convergent and make the above expressions somewhat suitable for a numerical implementation.
Let us point out, however, that more explicit formulas can be obtained if we consider integer values of $s$. In the next section we will focus our attention, in particular, to the case when
the argument $s$ assumes all integer negative values.

\section{$\zeta_{H}(s,i x)$ and $\zeta'_{H}(s,i x)$ for negative integers $s$}\label{sec3}

In order to consider the case when $s=-n$, with $n\in\mathbb{N}_{0}$, we will utilize the results obtained in section \ref{sec2} valid for $\Re(s)<1$. It is convenient to first prove two results concerning the functions $F(s,x)$ and $G(s,x)$. These
will be useful later on for the analysis of the first derivative of $\zeta_{H}(s,i x)$.
To this end we have the following:

\begin{lemma}\label{lemma1}
  Let $s=-n+\varepsilon$ with $n\in\mathbb{N}_{0}$ and $\varepsilon>0$. For $x\in(0,1)$,
  in the limit as $\varepsilon\to 0$ one has the behaviors
  \begin{equation}\label{11}
   \Re F(-n+\varepsilon,x)=\frac{(2\pi x)^{n}}{n!}\left(1+(-1)^{n}\right)(H_{n}-\ln 2\pi x)+2\sum_{k=0 \atop n\neq 2k}^{\infty}\frac{\zeta_{R}(1+n-2k)}{(2k)!}(2\pi x)^{2k}+O(\varepsilon)\;,
  \end{equation}
   \begin{equation}\label{11a}
   \Re G(-n+\varepsilon,x)=\frac{(2\pi x)^{n}}{n!}\left(1+(-1)^{n+1}\right)(H_{n}-\ln 2\pi x)+2\sum_{k=0 \atop n\neq 2k+1}^{\infty}\frac{\zeta_{R}(n-2k)}{(2k+1)!}(2\pi x)^{2k+1}+O(\varepsilon)\;,
  \end{equation}
  and
  \begin{equation}\label{11b}
    \Im F(-n+\varepsilon,x)=\Im G(-n+\varepsilon,x)=-\pi\frac{(2\pi x)^{n}}{n!}+O(\varepsilon)\;,
  \end{equation}
  where $H_{n}$ denote the harmonic numbers defined as
  \begin{equation}
    H_{n}=\sum_{k=1}^{n}\frac{1}{k}\;.
  \end{equation}
\end{lemma}
{\it Proof}. The proof of the above result is based on a direct computation. By utilizing the Taylor expansion \cite{gradshtein07}
\begin{equation}\label{12}
  \frac{\pi}{\tan\pi(\varepsilon-n)}=\frac{1}{\varepsilon}-\pi\sum_{k=1}^{\infty}\frac{2^{2k}|B_{2k}|}{(2k)!}(\pi\varepsilon)^{2k-1}\;,
\end{equation}
it is not difficult to obtain the following behavior for the first term of $\Re F(s,x)$ as $\varepsilon\to 0$
\begin{eqnarray}\label{13}
  \lefteqn{ \frac{\pi}{\tan\pi(\varepsilon-n)}\frac{(2\pi x)^{n-\varepsilon}}{\Gamma(1+n-\varepsilon)}=\frac{(2\pi x)^{n}}{n!}\left[\frac{1}{\varepsilon}+\Psi(n+1)-\ln 2\pi x\right]}\nonumber\\
   &&+\frac{(2\pi x)^{n}}{n!}\left[\ln^{2}2\pi x+\Psi^{2}(n+1)-2\Psi(n+1)\ln 2\pi x-\Psi^{\prime}(n+1)-\frac{2}{3}\pi^{2}\right]\varepsilon+O(\varepsilon^{2})\;,
\end{eqnarray}
where we have kept the term of order $\varepsilon$ because it will be used later in the analysis of the derivative.
The only term requiring special treatment in the series appearing in (\ref{7}) is the one for which the argument of the Riemann zeta function
approaches $1$. By isolating that particular term one has the expansion
\begin{equation}\label{14}
  \sum_{k=0}^{\infty}\frac{\zeta_{R}(1+n-k-\varepsilon)}{k!}(2\pi x)^{k}=-\frac{(2\pi x)^{n}}{n!}\left[\frac{1}{\varepsilon}+\gamma\right]+\sum_{k=0 \atop n\neq k}^{\infty}\frac{\zeta_{R}(1+n-k)}{k!}(2\pi x)^{k}+O(\varepsilon)\;,
\end{equation}
where $\gamma$ denotes the Euler-Mascheroni constant.
The last term to consider is the polylogarithmic function in (\ref{7}). By utilizing the representation (\ref{6}) the following result can be obtained  \cite{erdelyi53}
\begin{equation}\label{15}
  \textrm{Li}_{1+n-\varepsilon}\left(e^{-2\pi x}\right)=(-1)^{n}\frac{(2\pi x)^{n}}{n!}\left(H_{n}-\ln 2\pi x\right)+\sum_{k=0 \atop n\neq k}^{\infty}\frac{\zeta_{R}(1+n-k)}{k!}(-1)^{k}(2\pi x)^{k}+O(\varepsilon)\;.
\end{equation}
By substituting (\ref{13}), (\ref{14}) and (\ref{15}) into the expression (\ref{7a}) and by using the fact that \cite{gradshtein07} $\Psi(n+1)=-\gamma+H_{n}$, one obtains the claim (\ref{11}). The second claim, (\ref{11b}), can be proved by following the same procedure outlined above and by utilizing the relation (\ref{7b}).
The last two expressions can be easily derived by exploiting (\ref{7b}), setting $s=-n+\varepsilon$ in (\ref{7}) and by expanding around $\varepsilon=0$.

It is instructive to verify that the results of the above lemma indeed reproduce for all negative integer $s$ the following
well known relation involving the Bernoulli polynomials \cite{gradshtein07}
\begin{equation}\label{21}
  \zeta_{H}(-n,i x)=-\frac{B_{n+1}(i x)}{n+1}\;,\quad n\in\mathbb{N}_{0}\;.
\end{equation}

For this purpose, let us consider negative even integers, namely $n=2m$ with $m\in\mathbb{N}_{0}$. By applying the results of lemma \ref{lemma1} to the relation (\ref{8})
we have
\begin{equation}\label{16}
  \Re\zeta_{H}(-2m,ix)=\frac{(-1)^{m}}{2}x^{2m}\;.
\end{equation}
Lemma \ref{lemma1} with $n=2m$ applied to (\ref{8a})  gives, instead,
\begin{equation}\label{17}
\Im\zeta_{H}(-2m,ix)=(2m)!\sum_{j=0}^{m-1}\frac{(-1)^{j+1}B_{2(m-j)}}{[2(m-j)]!(2j+1)!}x^{2j+1}+\frac{(-1)^{m+1}}{2m+1}x^{2m+1}\;,
\end{equation}
with the understanding that for $m=0$ the first term in (\ref{17}) vanishes and where we have used the relation $\zeta_{R}(0)=-1/2$ and the following properties of the Riemann
zeta function valid for positive integers $m$ \cite{gradshtein07}
\begin{equation}\label{18}
  \zeta_{R}(2m)=(-1)^{m+1}\frac{2^{2m-1}\pi^{2m}B_{2m}}{(2m)!}\;,\qquad \zeta_{R}(-2m)=0\;.
\end{equation}

In a completely analogous way, we can obtain similar results for negative odd integers. In fact, by applying lemma \ref{lemma1} with $s=-2m-1$, $m\in\mathbb{N}_{0}$, and the remark in (\ref{18}), to (\ref{8}) and (\ref{8a}) we are led to the results
\begin{equation}\label{19}
  \Re\zeta_{H}(-2m-1,ix)=(2m+1)!\sum_{j=0}^{m}\frac{(-1)^{j+1}B_{2(m-j+1)}}{[2(m-j+1)]!(2j)!}x^{2j}+\frac{(-1)^{m}}{2(m+1)}x^{2m+2}
\end{equation}
and
\begin{equation}\label{20}
  \Im\zeta_{H}(-2m-1,ix)=\frac{(-1)^{m}}{2}x^{2m+1}\;.
\end{equation}

It is not very difficult to show that the expressions obtained in (\ref{16}), (\ref{17}), (\ref{19}) and (\ref{20}) are in complete agreement with the relation (\ref{21})
once its real and imaginary part are extracted.

Let us now turn our attention to the analysis of the first derivative of $\zeta_{H}(s,i x)$, in (\ref{22}) and (\ref{22a}), for negative integer values of $s$.
In order to study this case the following lemma will be useful:
\begin{lemma}\label{lemma2}
  Let $s=-n+\varepsilon$ with $n\in\mathbb{N}_{0}$ and $\varepsilon>0$.
  In the limit as $\varepsilon\to 0$ one has the behaviors
  \begin{eqnarray}\label{24}
   \lefteqn{\left(\Re F\right)'(-n+\varepsilon,x)=-\frac{\pi^{2}}{6}\frac{(2\pi x)^{n}}{n!}[2+(-1)^{n+1}]-2\sum_{k=0 \atop n\neq 2k}^{\infty}\frac{\zeta'_{R}(1+n-2k)}{(2k)!}(2\pi x)^{2k}}\nonumber\\
   &&+\frac{(2\pi x)^{n}}{2\,n!}[1+(-1)^{n}]\left[\ln^{2}2\pi x+\Psi^{2}(n+1)-2\Psi(n+1)\ln 2\pi x-\Psi^{\prime}(n+1)+2\gamma_{1}\right]+O(\varepsilon)\;,
  \end{eqnarray}
   \begin{eqnarray}\label{24a}
   \lefteqn{\left(\Re G\right)'(-n+\varepsilon,x)=-\frac{\pi^{2}}{6}\frac{(2\pi x)^{n}}{n!}[2+(-1)^{n}]-2\sum_{k=0 \atop n\neq 2k+1}^{\infty}\frac{\zeta'_{R}(n-2k)}{(2k+1)!}(2\pi x)^{2k+1}}\nonumber\\
   &&+\frac{(2\pi x)^{n}}{2\,n!}[1+(-1)^{n+1}]\left[\ln^{2}2\pi x+\Psi^{2}(n+1)-2\Psi(n+1)\ln 2\pi x-\Psi^{\prime}(n+1)+2\gamma_{1}\right]+O(\varepsilon)\;,
  \end{eqnarray}
  which hold for $x\in(0,1)$.
  \end{lemma}
{\it Proof}. The proof is based on the expansion of (\ref{9}) in the neighborhood of $s=-n$.  By exploiting the Laurent series in (\ref{13}) and \cite{gradshtein07}
\begin{equation}\label{25}
  \frac{2\pi}{\sin2\pi(\varepsilon-n)}=\frac{1}{\varepsilon}+2\pi\sum_{k=1}^{\infty}\frac{2(2^{2k-1}-1)|B_{2k}|}{(2k)!}(2\pi\varepsilon)^{2k-1}\;,
\end{equation}
one obtains, for the terms in the first line of (\ref{9}), the expansion
\begin{eqnarray}\label{26}
\lefteqn{-\frac{\pi}{\tan\pi (\varepsilon-n)}\frac{(2\pi x)^{n-\varepsilon}}{\Gamma(1+n-\varepsilon)}\left[\frac{2\pi}{\sin2\pi(\varepsilon-n)}+\ln2\pi x-\Psi(1+n-\varepsilon)\right]=-\frac{(2\pi x)^{n}}{n!}\frac{1}{\varepsilon^{2}}}\nonumber\\
&&+\frac{(2\pi x)^{n}}{2\,n!}\left[\ln^{2}2\pi x+\Psi^{2}(n+1)-2\Psi(n+1)\ln 2\pi x-\Psi^{\prime}(n+1)-\frac{2}{3}\pi^{2}\right]+O(\varepsilon)\;.
\end{eqnarray}
For the series containing the derivative of the Riemann zeta function in (\ref{9}) we have the behavior
\begin{equation}\label{27}
  -\sum_{k=0}^{\infty}\frac{\zeta'_{R}(1+n-k-\varepsilon)}{k!}(2\pi x)^{k}=\frac{(2\pi x)^{n}}{n!}\left[\frac{1}{\varepsilon^{2}}+\gamma_{1}\right]-\sum_{k=0 \atop n\neq k}^{\infty}\frac{\zeta_{R}'(1+n-k)}{k!}(2\pi x)^{k}\;,
\end{equation}
where $\gamma_{1}$ denotes the first Stieltjes constant. The last term to take into account is the derivative of the polylogarithmic function
in (\ref{9}). By differentiating the representation (\ref{6}) (with the sign of the exponent changed) one readily has
\begin{equation}\label{28}
  \textrm{Li}_{1-s}^{(1)}\left(e^{-2\pi x}\right)=\Gamma(s)(2\pi x)^{-s}\left[\ln 2\pi x-\Psi(s)\right]+\sum_{k=0}^{\infty}\frac{\zeta'_{R}(1-s-k)}{k!}(-1)^{k}(2\pi x)^{k}\;.
\end{equation}
In a neighborhood of $s=-n$, one can expand the first term of the previous expression to obtain
\begin{eqnarray}\label{29}
  \lefteqn{\Gamma(-n+\varepsilon)(2\pi x)^{n-\varepsilon}\left[\ln 2\pi x-\Psi(-n+\varepsilon)\right]=\frac{(-1)^{n}(2\pi x)^{n}}{n!}\frac{1}{\varepsilon^{2}}}\nonumber\\
  &&+\frac{(-1)^{n+1}(2\pi x)^{n}}{2\,n!}\left[\ln^{2}2\pi x+\Psi^{2}(n+1)-2\Psi(n+1)\ln 2\pi x-\Psi^{\prime}(n+1)+\frac{\pi^{2}}{3}\right]+O(\varepsilon)\;.
\end{eqnarray}
By exploiting the expansion for the derivative of the Riemann zeta function we finally get
\begin{eqnarray}\label{30}
  \textrm{Li}_{1+n-\varepsilon}^{(1)}\left(e^{-2\pi x}\right)&=&\frac{(-1)^{n+1}(2\pi x)^{n}}{2\,n!}\left[\ln^{2}2\pi x+\Psi^{2}(n+1)-2\Psi(n+1)\ln 2\pi x-\Psi^{\prime}(n+1)+2\gamma_{1}+\frac{\pi^{2}}{3}\right]\nonumber\\
  &+&\sum_{k=0 \atop n\neq k}^{\infty}\frac{\zeta'_{R}(1+n-k)}{k!}(-1)^{k}(2\pi x)^{k}+O(\varepsilon)\;.
\end{eqnarray}
By using the results obtained in (\ref{26}), (\ref{27}) and (\ref{30}) in the expression (\ref{9}) we arrive at the claim (\ref{24}).
The second claim, namely (\ref{24a}), can be easily proved along the same lines by noticing that
\begin{equation}\label{31}
  \Re G'(s,x)=\Re F'(s,x)+2\textrm{Li}^{(1)}_{1-s}\left(e^{-2\pi x}\right)\;.
\end{equation}

The above lemma allows us to compute the real and imaginary part of the derivative of the Hurwitz zeta function $\zeta_{H}(s,ix)$ for all
negative integers. Let us consider, first, the even integers $n=2m$ with $m\in\mathbb{N}_{0}$. From equation (\ref{22}) we can easily see that
terms proportional to $\sin(s\pi/2)$ do not contribute to $\Re\zeta_{H}(-2m,ix)$. By noticing that for all $m\in\mathbb{N}_{0}$
\begin{equation}\label{32}
  \Re F(-2m,ix)=\frac{(2\pi)^{2m+1}}{(2m)!}(-1)^{m}\left[\Im\zeta_{H}(-2m,ix)+2\textrm{Li}_{2m+1}\left(e^{-2\pi x}\right)\right]\;,
\end{equation}
which can be derived from (\ref{8a}) and (\ref{7b}), we arrive at the following expression
\begin{equation}\label{33}
  \Re\zeta'_{H}(-2m,ix)=\frac{\pi}{2}\Im \zeta_{H}(-2m,ix)+\frac{(-1)^{m+1}x^{2m}}{2}\ln x+(-1)^{m}\frac{(2m)!}{2(2\pi)^{2m}}\textrm{Li}_{2m+1}\left(e^{-2\pi x}\right)\;.
\end{equation}
In addition, by utilizing the result (\ref{24a}) of lemma \ref{lemma2} and the relation (\ref{32}) we obtain, for the imaginary part of the derivative of $\zeta_{H}(s,ix)$ at negative even integers, from (\ref{22a}), the formula
\begin{eqnarray}\label{34}
  \Im\zeta'_{H}(-2m,ix)&=&\Im \zeta_{H}(-2m,ix)\left(\ln 2\pi+\gamma-H_{2m}\right)+\frac{\pi}{2}(-1)^{m+1}x^{2m}\nonumber\\
  &+&(-1)^{m+1}\frac{(2m)!}{\pi(2\pi)^{2m}}\sum_{k=0}^{\infty}\frac{\zeta_{R}'(2m-2k)}{(2k+1)!}(2\pi x)^{2k+1}\;.
\end{eqnarray}

Let us discuss next the case of odd negative integers $s=-2m-1$, with $m\in\mathbb{N}_{0}$. First, we notice that by using (\ref{8})
we get the following relation valid for all $m\in\mathbb{N}_{0}$
\begin{equation}\label{35}
  \Re F(-2m-1,x)=\frac{(2\pi)^{2m+2}}{(2m+1)!}(-1)^{m+1}\Re\zeta_{H}(-2m-1,ix)\;.
\end{equation}
The result (\ref{24}) together with the remark (\ref{35}) leads us to the expression
\begin{eqnarray}\label{36}
  \Re\zeta'_{H}(-2m-1,ix)&=&\Re \zeta_{H}(-2m-1,ix)\left(\ln 2\pi+\gamma-H_{2m+1}\right)+\frac{\pi}{2}(-1)^{m}x^{2m+1}\nonumber\\
  &+&(-1)^{m}\frac{(2m+1)!}{\pi(2\pi)^{2m+1}}\sum_{k=0}^{\infty}\frac{\zeta_{R}'(2m-2k+2)}{(2k)!}(2\pi x)^{2k}\;.
\end{eqnarray}
A similar result for the imaginary part at $s=-2m-1$ can be found by noticing that the terms proportional to $\cos(s\pi/2)$ do not contribute
and that for $m\in\mathbb{N}_{0}$
\begin{equation}\label{37}
  \Re G(-2m-1,x)=\frac{(2\pi)^{2m+2}}{(2m+1)!}(-1)^{m+1}\Re\zeta_{H}(-2m-1,ix)-2\textrm{Li}_{2m+2}\left(e^{-2\pi x}\right)\;,
\end{equation}
which is obtained from (\ref{7b}) and (\ref{8}). The last remarks allow us to derive
\begin{equation}\label{38}
  \Im\zeta'_{H}(-2m-1,ix)=-\frac{\pi}{2}\Re\zeta_{H}(-2m-1,ix)+\frac{(-1)^{m+1}x^{2m+1}}{2}\ln x+(-1)^{m}\frac{(2m+1)!}{2(2\pi)^{2m+1}}\textrm{Li}_{2m+2}\left(e^{-2\pi x}\right)\;.
\end{equation}

It is important to point out that the results in (\ref{33}) and in (\ref{38}) are in complete agreement with the values of $\Re\zeta'_{H}(-2m,ix)$
and $\Im\zeta'_{H}(-2m-1,ix)$ obtained, by following a different procedure, in the paper \cite{adesi02}.

\section{$\zeta_{H}(s,i x)$ and $\zeta'_{H}(s,i x)$ for positive integers $s$}\label{sec4}

For the analysis of the real and imaginary part of $\zeta_{H}(s,ix)$ for positive integer values of $s$, it is convenient to prove the following:
 \begin{lemma}\label{lemma3}
  Let $s=n+\varepsilon$ with $n\in\mathbb{N}$ and $\varepsilon>0$. Then as $\varepsilon\to 0$, we have
  \begin{eqnarray}\label{40a}
  F'(n+\varepsilon,x)&=&(-1)^{n+1}i\pi\,(n-1)!(2\pi x)^{-n}-2\sum_{k=0}^{\infty}\frac{\zeta_{R}'(1-n-2k)}{(2k)!}(2\pi x)^{2k}\nonumber\\
  &-&\left[1+(-1)^{n}\right](n-1)!(2\pi x)^{-n}\left(\ln 2\pi x-\Psi(n)\right)+O(\varepsilon)\;,
  \end{eqnarray}
  \begin{eqnarray}\label{40b}
  G'(n+\varepsilon,x)&=&(-1)^{n+1}i\pi\,(n-1)!(2\pi x)^{-n}-2\sum_{k=0}^{\infty}\frac{\zeta_{R}'(-n-2k)}{(2k+1)!}(2\pi x)^{2k+1}\nonumber\\
  &+&\left[1+(-1)^{n+1}\right](n-1)!(2\pi x)^{-n}\left(\ln 2\pi x-\Psi(n)\right)+O(\varepsilon)\;,
   \end{eqnarray}
  which hold for $x\in(0,1)$.
  \end{lemma}
{\it Proof}. By differentiating (\ref{6}) with respect to the variable $s$ and by subsequently setting $s=n$
we obtain the following expression
\begin{equation}\label{41}
  \textrm{Li}^{(1)}_{1-n}\left(e^{2\pi x}\right)=(-1)^{n}(n-1)!(2\pi x)^{-n}\left[i\pi+\ln 2\pi x-\Psi(n)\right]+\sum_{k=0}^{\infty}\frac{\zeta'_{R}(1-n-k)}{k!}(2\pi x)^{k}\;.
\end{equation}
In addition, directly from equation (\ref{28}), we have
\begin{equation}\label{42}
  \textrm{Li}_{1-n}^{(1)}\left(e^{-2\pi x}\right)=(n-1)!(2\pi x)^{-n}\left[\ln 2\pi x-\Psi(n)\right]+\sum_{k=0}^{\infty}\frac{\zeta'_{R}(1-n-k)}{k!}(-1)^{k}(2\pi x)^{k}\;.
\end{equation}
From the results in (\ref{41}) and (\ref{42}) it is straightforward to prove the claims (\ref{40a}) and (\ref{40b}) by using the relations
\begin{equation}\label{43}
  F'(s,x)=-\textrm{Li}^{(1)}_{1-s}(e^{2\pi x})-\textrm{Li}^{(1)}_{1-s}(e^{-2\pi x})\;,\quad G'(s,x)=-\textrm{Li}^{(1)}_{1-s}(e^{2\pi x})+\textrm{Li}^{(1)}_{1-s}(e^{-2\pi x})\;,
\end{equation}
which are obtained by differentiation of (\ref{5a}).

Let us consider even positive integers $s=2m$ with $m\in\mathbb{N}$. For $\varepsilon\to 0$, we have the following expansion for the terms proportional to $F(s,x)$ and $G(s,x)$ in (\ref{5})
\begin{equation}\label{44}
  \frac{\Gamma(1-2m-\varepsilon)}{(2\pi)^{1-2m-\varepsilon}}\sin\left(\frac{\pi}{2}(2m+\varepsilon)\right)=\frac{\pi}{2}\frac{(-1)^{m}(2\pi)^{2m-1}}{(2m-1)!}+O(\varepsilon)\;,
\end{equation}
and
\begin{equation}\label{45}
  \frac{\Gamma(1-2m-\varepsilon)}{(2\pi)^{1-2m-\varepsilon}}\cos\left(\frac{\pi}{2}(2m+\varepsilon)\right)=\frac{(-1)^{m}(2\pi)^{2m-1}}{(2m-1)!}\left[\frac{1}{\varepsilon}+\ln 2\pi+\gamma-H_{2m-1}\right]+O(\varepsilon)\;.
\end{equation}
By noticing that for $s=2m$ the relation (\ref{4}) quickly leads to $G(2m,x)=0$, the last two results, together with (\ref{5}), provide us with the formula
\begin{equation}\label{46}
  \zeta_{H}(2m,ix)=\frac{\pi}{2}\frac{(-1)^{m}(2\pi)^{2m-1}}{(2m-1)!}\left[F(2m,x)+i\frac{2}{\pi}G'(2m,x)\right]\;.
\end{equation}
From the last expression, we can easily compute the real and imaginary part of $\zeta_{H}(2m,ix)$ by making use of (\ref{40b}) in lemma \ref{lemma3}.
More explicitly, we obtain, for $0<x<1$,
\begin{equation}\label{47}
  \Re\zeta_{H}(2m,ix)=(-1)^{m}\frac{(2\pi)^{2m}}{4(2m-1)!}F(2m,x)+\frac{(-1)^{m}}{2}x^{-2m}\;,
\end{equation}
and
\begin{equation}\label{48}
  \Im\zeta_{H}(2m,ix)=(-1)^{m+1}\frac{(2\pi)^{2m}}{\pi(2m-1)!}\sum_{k=0}^{\infty}\frac{\zeta'_{R}(-2(m+k))}{(2k+1)!}(2\pi x)^{2k+1}\;.
\end{equation}
We would like to point out that when $s=n$ with $n\in\mathbb{N}$, $F(s,x)$ and $G(s,x)$ can be written, for $0<w<1$, in terms of
elementary functions due to the following relation enjoyed by the polylogarithmic function \cite{lewin81}
\begin{equation}\label{49}
  \textrm{Li}_{-n}\left(w\right)=\frac{1}{(1-w)^{n+1}}\sum_{k=0}^{n-1}\left<n\atop k\right>w^{n-k}\;,
\end{equation}
where the numerical coefficients appearing in the sum are the Eulerian numbers defined as \cite{carlitz59}
\begin{equation}
  \left<n\atop k\right>=\sum_{j=0}^{k+1}(-1)^{j} {n+1\choose j}(k-j+1)^{n}\;.
\end{equation}

We can proceed in a similar way in order analyze the case when $s=2m+1$ with $m\in\mathbb{N}$, without considering $s=1$ where the Hurwitz zeta function presents a simple pole with residue one. In this situation
we have the expansions
\begin{equation}\label{50}
  \frac{\Gamma(-2m-\varepsilon)}{(2\pi)^{-2m-\varepsilon}}\sin\left(\frac{\pi}{2}(2m+1+\varepsilon)\right)=\frac{(-1)^{m+1}(2\pi)^{2m}}{(2m)!}\left[\frac{1}{\varepsilon}+\ln 2\pi+\gamma-H_{2m}\right]+O(\varepsilon)
\end{equation}
and
\begin{equation}\label{51}
  \frac{\Gamma(-2m-\varepsilon)}{(2\pi)^{-2m-\varepsilon}}\cos\left(\frac{\pi}{2}(2m+1+\varepsilon)\right)=\frac{\pi}{2}\frac{(-1)^{m}(2\pi)^{2m}}{(2m)!}+O(\varepsilon)\;,
\end{equation}
which, supplemented with the condition $F(2m+1,x)=0$ for $m\in\mathbb{N}$, allow us to obtain, from (\ref{5}), the formula
\begin{equation}\label{52}
  \zeta_{H}(2m+1,ix)=\frac{\pi}{2}\frac{(-1)^{m}(2\pi)^{2m}}{(2m)!}\left[iG(2m+1,x)-\frac{2}{\pi}F'(2m+1,x)\right]\;.
\end{equation}
By exploiting the result (\ref{40a}) of lemma \ref{lemma3} to evaluate $F'(2m+1,x)$, it is not very difficult to extract the real and imaginary part of (\ref{52}).
In more detail one has, for $0<x<1$,
\begin{equation}\label{53}
  \Re\zeta_{H}(2m+1,ix)=(-1)^{m}\frac{2(2\pi)^{2m}}{(2m)!}\sum_{k=0}^{\infty}\frac{\zeta'_{R}(-2(m+k))}{(2k)!}(2\pi x)^{2k}\;,
\end{equation}
and
\begin{equation}\label{54}
  \Im\zeta_{H}(2m+1,ix)=(-1)^{m}\frac{(2\pi)^{2m+1}}{4(2m)!}G(2m+1,x)+\frac{(-1)^{m+1}}{2}x^{-2m-1}\;.
\end{equation}

Let us focus next on the study of the real and imaginary part of the derivative of $\zeta_{H}(s,ix)$. For this purpose, it is useful
to prove the following result:
\begin{lemma}\label{lemma4}
  Let $s=n+\varepsilon$ with $n\in\mathbb{N}$ and $\varepsilon>0$. Then as $\varepsilon\to 0$, we have
  \begin{eqnarray}\label{55a}
  F''(n+\varepsilon,x)&=&(-1)^{n+1}\pi^{2}(n-1)!(2\pi x)^{-n}+2\sum_{k=0}^{\infty}\frac{\zeta_{R}''(1-n-2k)}{(2k)!}(2\pi x)^{2k}\nonumber\\
  &+&\left[1+(-1)^{n}\right](n-1)!(2\pi x)^{-n}\left(\ln^{2}2\pi x-2\Psi(n)\ln 2\pi x-\Psi^{2}(n)+\Psi'(n)\right)\nonumber\\
  &+&(-1)^{n}2\pi i\,(n-1)!(2\pi x)^{-n}\left(\ln 2\pi x-\Psi(n)\right)+O(\varepsilon)\;,
  \end{eqnarray}
  \begin{eqnarray}\label{55b}
  G''(n+\varepsilon,x)&=&(-1)^{n+1}\pi^{2}(n-1)!(2\pi x)^{-n}+2\sum_{k=0}^{\infty}\frac{\zeta_{R}''(-n-2k)}{(2k+1)!}(2\pi x)^{2k+1}\nonumber\\
  &-&\left[1+(-1)^{n+1}\right](n-1)!(2\pi x)^{-n}\left(\ln^{2}2\pi x-2\Psi(n)\ln 2\pi x-\Psi^{2}(n)+\Psi'(n)\right)\nonumber\\
  &+&(-1)^{n}2\pi i\,(n-1)!(2\pi x)^{-n}\left(\ln 2\pi x-\Psi(n)\right)+O(\varepsilon)\;,
   \end{eqnarray}
  which hold for $x\in(0,1)$.
  \end{lemma}
{\it Proof}. By differentiating (\ref{6}) twice with respect to $s$ and by then setting $s=n$ we arrive at the expression
\begin{eqnarray}\label{56}
  \textrm{Li}^{(2)}_{1-n}\left(e^{2\pi x}\right)&=&(-1)^{n+1}\pi^{2}(n-1)!(2\pi x)^{-n}+\sum_{k=0}^{\infty}\frac{\zeta''_{R}(1-n-k)}{k!}(2\pi x)^{k}\nonumber\\
  &+&(-1)^{n}2\pi i\,(n-1)!(2\pi x)^{-n}(\ln 2\pi x-\Psi(n))\nonumber\\
  &+&(-1)^{n}(n-1)!(2\pi x)^{-n}\left[\ln^{2}2\pi x-2\Psi(n)\ln 2\pi x+\Psi^{2}(n)+\Psi'(n)\right]\;.
\end{eqnarray}
Analogously, the first derivative of (\ref{28}) leads, for positive integer values of $s$, to
\begin{eqnarray}\label{57}
   \textrm{Li}^{(2)}_{1-n}\left(e^{-2\pi x}\right)&=&(n-1)!(2\pi x)^{-n}\left[\ln^{2}2\pi x-2\Psi(n)\ln 2\pi x+\Psi^{2}(n)+\Psi'(n)\right]\nonumber\\
   &+&\sum_{k=0}^{\infty}\frac{\zeta''_{R}(1-n-k)}{k!}(-2\pi x)^{k}\;.
\end{eqnarray}
The last two expressions for the second derivative of the polylogarithmic function are sufficient in order to arrive at the claims (\ref{55a}) and (\ref{55b})
once we use them in the equations
\begin{equation}\label{58}
  F''(s,x)=\textrm{Li}^{(2)}_{1-s}(e^{2\pi x})+\textrm{Li}^{(2)}_{1-s}(e^{-2\pi x})\;,\quad G''(s,x)=\textrm{Li}^{(2)}_{1-s}(e^{2\pi x})-\textrm{Li}^{(2)}_{1-s}(e^{-2\pi x})\;,
\end{equation}
which are obtained from (\ref{43}) by differentiation.

Let us begin, once again, with the analysis of the case of even integers $s=2m$, $m\in\mathbb{N}$. By making use of the expansions in (\ref{44}) and (\ref{45}) in the
expressions (\ref{23}) for $\zeta'_{H}(s,ix)$, and by noticing that $G(2m,x)=0$ for $m\in\mathbb{N}$, we have
\begin{eqnarray}\label{59}
  \zeta'_{H}(2m,ix)&=&\frac{\pi}{2}\frac{(-1)^{m}(2\pi)^{2m-1}}{(2m-1)!}\left[\left(\ln 2\pi-\Psi(2m)\right)F(2m,x)+F'(2m,x)\right]\nonumber\\
  &-&i\frac{(-1)^{m+1}(2\pi)^{2m-1}}{(2m-1)!}\left[\left(\ln 2\pi-\Psi(2m)\right)G'(2m,x)+\frac{1}{2}G''(2m,x)\right]\;.
\end{eqnarray}
In order to obtain an explicit expression for $F'(2m,x)$, $G'(2m,x)$ and $G''(2m,x)$ we use (\ref{40a}) and (\ref{40b}) from lemma \ref{lemma3} and the result (\ref{55b}) from lemma \ref{lemma4}. By proceeding in this fashion and by using the relation (\ref{47}) we obtain the following result for the real part valid for $0<x<1$
\begin{eqnarray}\label{60}
  \Re\zeta'_{H}(2m,ix)&=&\Re\zeta_{H}(2m,ix)(\ln 2\pi+\gamma-H_{2m-1})+(-1)^{m+1}x^{-2m}(\ln 2\pi x+\gamma-H_{2m-1})\nonumber\\
  &+&(-1)^{m+1}\frac{(2\pi)^{2m}}{2(2m-1)!}\sum_{k=0}^{\infty}\frac{\zeta'_{R}(1-2(m+k))}{(2k)!}(2\pi x)^{2k}\;.
\end{eqnarray}
In a similar way, by using the formula (\ref{48}), we have for the imaginary part
\begin{eqnarray}\label{61}
  \Im\zeta'_{H}(2m,ix)&=&\Im\zeta_{H}(2m,ix)(\ln 2\pi+\gamma-H_{2m-1})+\frac{\pi(-1)^{m+1}}{2}x^{-2m}\nonumber\\
  &+&(-1)^{m}\frac{(2\pi)^{2m-1}}{(2m-1)!}\sum_{k=0}^{\infty}\frac{\zeta''_{R}(-2(m+k))}{(2k+1)!}(2\pi x)^{2k+1}\;,
\end{eqnarray}
with the condition $0<x<1$.

To conclude the analysis, let us consider the case when $s$ is a positive odd integer, namely $s=2m+1$, with $m\in\mathbb{N}$. The expansions
(\ref{50}) and (\ref{51}) employed in (\ref{23}) allow us to write
\begin{eqnarray}\label{62}
  \zeta'_{H}(2m+1,ix)&=&i\frac{\pi}{2}\frac{(-1)^{m}(2\pi)^{2m}}{(2m)!}\left[\left(\ln 2\pi-\Psi(2m+1)\right)G(2m+1,x)+G'(2m+1,x)\right]\nonumber\\
  &+&\frac{(-1)^{m}(2\pi)^{2m}}{(2m)!}\left[\left(\ln 2\pi-\Psi(2m+1)\right)F'(2m,x)+\frac{1}{2}F''(2m,x)\right]\;,
\end{eqnarray}
where we have used the fact that $F(2m+1,x)=0$. The relations (\ref{40a}), (\ref{40b}) and (\ref{55b}) substituted in (\ref{62}) allow us to extract its real and imaginary part.
In fact, by exploiting (\ref{53}) we get the following expression valid for $0<x<1$
\begin{eqnarray}\label{63}
\Re\zeta'_{H}(2m+1,ix)&=&\Re\zeta_{H}(2m+1,ix)(\ln 2\pi+\gamma-H_{2m})+\frac{\pi(-1)^{m+1}}{2}x^{-2m-1}\nonumber\\
&+&(-1)^{m+1}\frac{(2\pi)^{2m}}{(2m)!}\sum_{k=0}^{\infty}\frac{\zeta''_{H}(-2(m+k))}{(2k)!}(2\pi x)^{2k}\;.
\end{eqnarray}
In a similar way, the use of the relation (\ref{54}) provides us with a formula for the imaginary part of (\ref{62}). In more detail one has, for $0<x<1$,
\begin{eqnarray}\label{64}
  \Im\zeta'_{H}(2m+1,ix)&=&\Im\zeta_{H}(2m+1,ix)(\ln 2\pi+\gamma-H_{2m})+(-1)^{m}x^{-2m-1}(\ln 2\pi x+\gamma-H_{2m})\nonumber\\
  &+&(-1)^{m+1}\frac{\pi(2\pi)^{2m}}{(2m)!}\sum_{k=0}^{\infty}\frac{\zeta'_{R}(-1-2(m+k))}{(2k+1)!}(2\pi x)^{2k+1}\;.
\end{eqnarray}

\section{Concluding Remarks}

In this work we have utilized Jonqui\`{e}re's representation of the Hurwitz zeta function in order to find expressions for the real and
imaginary part of $\zeta_{H}(s,ix)$ and its first derivative. We have then specialized the obtained results to the case of integer $s$, namely $s\in\mathbb{Z}\backslash\{1\}$,
where explicit formulas, which involve polylogarithmic functions and the Riemann zeta function, have been presented. The expressions that we have found can be
directly applied to the computation of the production rate of particles and anti-particles in strong electric fields in the setting of a higher-dimensional Minkowski spacetime (see e.g. \cite{blau91}). A further application of the results obtained in this work can be found in the analysis of the one-loop partition function and Casimir energy for scalar fields at finite temperature and chemical potential. The high temperature expansion of these quantities explicitly depends on the Hurwitz zeta function of imaginary second argument evaluated at integer points (see e.g. \cite{dowker89,kirsten91,kirsten91a}).

In general, these results could find applications to the analysis of the Schwinger mechanism in more general settings. This would include cases when the relevant one-loop effective action contains the Hurwitz zeta function of imaginary second argument and its derivative evaluated at specific integer points.
More specifically, it might be possible to apply the formulas obtained in this work to the study of the Schwinger mechanism on product manifolds and also
to the analysis of thermal corrections to the pair production rate (since the relevant manifold in this case would be of the type $M\times S^{1}$, a particular case of product manifold).
The evaluation of finite temperature corrections to the Schwinger pair production rate is a subject of particular interest since the results that one obtains depend on the specific formalism used (see for instance \cite{kim09} and references therein). The results presented here can be used, in the framework of zeta function regularization, in order to provide a way to compute the thermal corrections which is different from the ones find in the literature.
Obviously these claims need to be verified and deserve further investigation.

We would like to make a final remark of mathematical character. The expressions obtained for
the real and imaginary part of $\zeta_{H}(s,ix)$ for $s\in\mathbb{Z}\backslash\{1\}$ and its first derivative actually provide summation formulas for specific series involving the first and second derivative of the Riemann zeta function. Based on this observation, it seems worth to study this point in more detail in order to understand whether the methods used here can provide new summation formulas for series involving first and second derivative of the Riemann zeta function.

\end{document}